\documentclass[english]{article}
\usepackage[T1]{fontenc}
\usepackage[latin9]{inputenc}
\usepackage{textcomp}
\usepackage{amssymb}
\usepackage{stmaryrd}

\makeatletter

\let\SF@@footnote\footnote
\def\footnote{\ifx\protect\@typeset@protect
    \expandafter\SF@@footnote
  \else
    \expandafter\SF@gobble@opt
  \fi
}
\expandafter\def\csname SF@gobble@opt \endcsname{\@ifnextchar[
  \SF@gobble@twobracket
  \@gobble
}
\edef\SF@gobble@opt{\noexpand\protect
  \expandafter\noexpand\csname SF@gobble@opt \endcsname}
\def\SF@gobble@twobracket[#1]#2{}

\@ifundefined{date}{}{\date{}}
\makeatother

\usepackage{babel}
\begin{document}
\title{Inferences and Modal Vocabulary\\
{\normalsize{}Logical Understanding II}}
\author{Florian Richter}
\maketitle
\begin{abstract}
Deduction is the one of the major forms of inferences and commonly
used in formal logic. This kind of inference has the feature of monotonicity,
which can be problematic. There are different types of inferences
that are not monotonic, e.g. abductive inferences. The debate between
advocates and critics of abduction as a useful instrument can be reconstructed
along the issue, how an abductive inference warrants to pick out one
hypothesis as the best one. But how can the goodness of an inference
be assessed? Material inferences express good inferences based on
the principle of material incompatibility. Material inferences are
based on modal vocabulary, which enriches the logical expressivity
of the inferential relations. This leads also to certain limits in
the application of labeling in machine learning. I propose a modal
interpretation of implications to express conceptual relations.
\end{abstract}

\section{Introduction\protect\footnote{This paper is based on a talk that I gave at the International Conference
on Information and Computer Technologies (ICICT 2020) in San Jose.}}

\subsection{Context \textendash{} Commitments}

To contextualize this paper, I will start with two commitments. (1.)
I believe that philosophy can contribute to the debates in the field
of artificial intelligence. The philosophy of language and the philosophy
of logic provide us with many tools and insights of language use and
how we use concepts. 

(2.) I believe that inferential relations govern our concept use.
So, if we label something, what we actually do, is drawing an inference.\footnote{Charles Sanders Peirce thinks that a perceptual judgment is really
an abductive inference, that can be made explicit.\cite{Peirce1935}
(CP 5.3) In epistemological debates in philosophy it is an extensively
discussed topic, whether to perceive something is to draw an inference
or not. } There are different kinds of inferential classifications: to label
something or to describe something are actions that are governed by
inferential relations.\cite{Brandom2009} It could also be described
as a difference between classificatory and conceptual inferences.\cite{Stekeler-Weithofer2015} 

\subsection{Problem}

There are different kinds of inferences, like e.g. induction, deduction,
and abduction. \textendash{} Deduction is the one of the major forms
of inferences and commonly used in formal logic. This kind of inference
has the feature of monotonicity, which can be problematic as the famous
example of the ``flying penguin'' shows.\cite{Ertel2017} Take the
premises: (1.) Tweety is a penguin. (2.) Penguins are birds. (3.)
Birds can fly. From these premises can be concluded, that Tweety can
fly. If we add the statement that penguins cannot fly, we can draw
the consequence that Tweety can\textasciiacute t fly, but the former
statement that Tweety can fly is still valid. So, we have an inconsistent
knowledge base.\cite{Ertel2017} \textendash{} And nothing is more
awful for a logician, if there is an inconsistency. It is possible
to construct a consistent metalanguage like a calculus for propositional
logic with the introduction of deductive reasoning (\emph{modus ponens})
and the feature of monotonicity, but the example above shows the problematic
point of using deductive reasoning for natural language use and it
is mainly connected with the feature of monotonicity, which therefore
has to be omitted. 

\section{Nonmonotic Inferences }

There are different types of inferences that are not monotonic, e.g.
abductive inferences do not require an exhaustive list of conditions
or premises that might be in the end inconsistent, because it is more
about proposing premises as hypotheses. Charles Sanders Peirce writes:
``Deduction proves that something \textbf{must be}; Induction shows
that something \textbf{actually is} operative; Abduction merely suggests
that something \textbf{may be}.''\cite{Peirce1935} (CP 5.171) The
debate between advocates and critics of abduction as a useful instrument
can be reconstructed along the issue, how an abductive inference warrants
to pick out one hypothesis as the best one (given that there is an
already established set of hypotheses to choose from).\cite{Douven2017}
There are approaches that claim that the best hypothesis is picked
out by applying principles like coherence or simplicity to the explanations.\cite{Thagard1978}
It needs to be shown, how these principles can be at least spelled
out more concretely. 

\subsection{Abduction - Peirce}

Peirce worked extensively on this kind of inference, although he admitted
in several occasions that he confused induction and abduction. (He
called abduction hypothetic inference, but understood later that it
is actually an induction from qualities and that abduction is something
else.)\footnote{``By hypothetic inference, I mean, {[}\dots {]} an induction from
qualities''\cite{Peirce1935a} (CP 6.145) ``Only in almost everything
I printed before the beginning of this century I more or less mixed
up Hypothesis and Induction...''\cite{Peirce1958} (CP 8.227)} Peirce describes the form of the inference in the following way:
\begin{quote}
``Long before I first classed abduction as an inference it was recognized
by logicians that the operation of adopting an explanatory hypothesis
\textendash{} which is just what abduction is \textendash{} was subject
to certain conditions. Namely, the hypothesis cannot be admitted,
even as a hypothesis, unless it be supposed that it would account
for the facts or some of them. The form of inference, therefore, is
this:

The surprising fact, C, is observed;

But if A were true, C would be a matter of course,

Hence, there is reason to suspect that A is true.

Thus, A cannot be abductively inferred, or if you prefer the expression,
cannot be abductively conjectured until its entire content is already
present in the premiss, \textquoteleft If A were true, C would be
a matter of course.'{}''\cite{Peirce1935} (CP 5.188/189)
\end{quote}
To admit or propose a hypothesis as a hypothesis \emph{is only possible
with respect to the fact}. The inference presupposes the relation
of them in the premise. It is in fact not an act of cognition \textendash{}
of grasping the propositional content, that the premise expresses
\textendash , but an act of recognition of the relation.

In the end, what is important is to know which conditions should be
fulfilled that the abductive inference is a good inference. According
to him the ``question of the goodness of anything is whether that
thing fulfills its end.'' And the end of the abductive inference
is ``to lead to the avoidance of all surprise and to the establishment
of a habit of positive expectations that shall not be disappointed''.
One needs to be capable to experimentally verify the expectations
of the hypothesis. Peirce sees of course that this leads to the question:
``What are we to understand by experimental verification?'' \textendash{}
But experimental verification is governed by other principles, rules,
and inferences and then we are talking about the ``logic of induction''.\footnote{``Admitting, then, that the question of Pragmatism is the question
of Abduction, let us consider it under that form. What is good abduction?
What should an explanatory hypothesis be to be worthy to rank as a
hypothesis? Of course, it must explain the facts. But what other conditions
ought it to fulfill to be good? The question of the goodness of anything
is whether that thing fulfills its end. What, then, is the end of
an explanatory hypothesis? Its end is, through subjection to the test
of experiment, to lead to the avoidance of all surprise and to the
establishment of a habit of positive expectation that shall not be
disappointed. Any hypothesis, therefore, may be admissible, in the
absence of any special reasons to the contrary, provided it be capable
of experimental verification, and only insofar as it is capable of
such verification. This is approximately the doctrine of pragmatism.
But just here a broad question opens out before us. What are we to
understand by experimental verification? The answer to that involves
the whole logic of induction.''\cite{Peirce1935} (CP 5.197)}

\subsection{Abduction - Best Explanation}

There are other suggestions for the goodness of an abductive inference
(without diving in the vast discourse in the philosophy of science).
Paul Thagard claims that a theory explains the facts better, if it
does not make additional statements that might not seem plausible
or hard to combine with other explanations, facts, or theories. He
mentions the advantages of Charles Darwin\textasciiacute s theory
of natural selection to explain certain facts, like e.g. the geographic
distribution of species. Another example that the mentions is the
``phlogiston theory'' that ``burning objects \emph{give off} the
substance phlogiston, whereas, according to Lavoisier, burning objects
\emph{combine} with oxygen. The main point of Lavoisier\textasciiacute s
argument is that his theory can explain the fact that bodies undergoing
combustion increase in weight rather than decrease {[}...{]}. To explain
the same fact, proponents of the phlogiston theory had to make such
odd assumptions as that the phlogiston that was supposedly given off
had \textquoteleft negative weight.'{}''\cite{Thagard1978} (77/78)
Thagard suggests therefore that the ``above arguments exemplify three
important criteria for determining the best explanation.'' These
criteria are: ``\emph{consilience}, \emph{simplicity}, and \emph{analogy}.''
Important is here to mention that by ``criteria'' he does ``not
mean necessary or sufficient conditions.''\cite{Thagard1978} (77/78)

It is also argued that the hypothesis needs to yield an explanation
that is sufficiently good enough\cite{Lipton1993}, but it is not
clear what ``sufficiently good enough'' might mean and involve.
Another example of nonmontonic inferences is clearer in this respect:
material inferences. They express good inferences based on the principle
of material incompatibility. The claim, if p then q, is incompatible
with the claim that it is possible that p and not-q.\cite{Brandom2008}
Instead of formal inconsistencies (like in the case of deduction)
a kind of material incoherence can be expressed and is taken as a
principle to distinguish good inferences from bad inferences, but
the same problem arises, because if it is not possible to deal with
inconsistent predicates or propositions like in the example of the
flying penguin, it would be also difficult to deal with incoherent
or incompatible ones. 

\subsection{Material Inferences}

Brandom develops an incompatibility semantics in his book \emph{Between
Saying and Doing}. Brandom uses the deontic vocabulary ``commitment''
and ``entitlement''. Being committed to a claim does not only mean
to state that the claim is true, but also to commit yourself to be
able to justify the claim. It is your responsibility to give a reason
for your commitment and if you have a reason you might be also entitled
to the claim. If you are e.g. committed to a plane figure being rectangular,
you also have to be committed to its being polygonal. ``And the old
nautical meteorological homily, \textquoteleft Red sky at night, sailor\textasciiacute s
delight; red sky in morning, sailor take warning,' tells us that anyone
who sees a colorful sunrise is entitled to the claim that a storm
that day is probable.'' Brandom writes that this ``reasoning is
only probative, not dispositive'', because the ``colorful sunrise
provides \emph{some} reason to predict a storm, but does not yet settle
the matter. Other considerations, such as a rising barometer, may
license one not to draw the conclusion one would otherwise be entitled
to by the original evidence.''\cite{Brandom2008} (119/120)

A deductive inferential relation means that if you are committed to
one claim you also \emph{have to} be committed to the consequence
of the claim. A rectangular figure is also polygonal. Both predicates
are not incompatible with each other. An inductive inferential relation
means that one who is entitled to one claim is also entitled to the
the consequence of the claim. Important is here that the ``reasoning
is only probative''. Brandom talks here about predictions. In machine
learning one of the interests lies in making predictions from experience
or data. If you commit yourself to the claim that ``Pedro is a donkey'',
then you have to be committed to the claim that ``Pedro is a mammal.''
All that is incompatible with the second claim is also incompatible
with the first one.\footnote{``Thus \textquoteleft Pedro is a donkey,' incompatibility-entails
\textquoteleft Pedro is a mammal,' for everything incompatible with
Pedro's being a mammal (for instance, Pedro's being an invertebrate,
an electronic apparatus, a prime number) is incompatible with Pedro's
being a donkey.'' \cite{Brandom2008} (121)}

\section{Incompatibility}

The inferential relations of deduction and induction can be made explicit
by the deontic vocabulary of ``commitment'' and ``entitlement''.
For Brandom it seems possible to make ``\emph{incompatibility}-entailments''
explicit by a specific ``kind of inferences'': ``\emph{modally
qualified conditionals}''. He writes that if ``two \emph{properties}
(such as being a mammal and being an invertebrate) are incompatible
then it is \emph{impossible} for any object simultaneously to exhibit
both.'' This means therefore ``it is \emph{impossible} for anything
to be a donkey and not be a mammal. That is why the incompatibility-entailment
in question supports counterfactuals such as \textquoteleft If my
first pet (in fact, let us suppose, a fish) \emph{had been} a donkey,
it \emph{would have been} a mammal.' We could say: \textquoteleft Necessarily,
anything that is a donkey is a mammal.'{}''\cite{Brandom2008} (121/122)

The modal vocabulary is a specific tool of expressing inferential
relations. Relations between species and genus can be made explicit
by this ``\emph{modally qualified conditionals}''. If a genus has
necessarily a certain characteristic, also the species has to have
it. There are also certain characteristics that do not necessarily
belong to the genus, because not all the species have it. Every mammal
has necessarily a heart, but not every mammal has necessarily legs
(e.g. whales or dolphins). This is connected with the semantic dimension
of the concepts. The essential idea of Brandom is that \emph{incompatibility}
does not have to be mistaken with \emph{inconsistency} on the formal,
logical side. It can be that one claim is not incompatible with other
claims individually, but it can be that it is incompatible with a
set of these individual claims, e.g. ``the claim that the piece of
fruit in my hand is a blackberry is incompatible with the \emph{two}
claims that it is red and that it is ripe, though not with either
individually''.\cite{Brandom2008} (123)

It is an important question, how incompatibility can be formlized.
Maybe one can formalize it by introducing an operator for incompatibility,
but then it has to be shown, how the this operator is primitive or
how it can be formulated by definition with other primitive operators.
Brandom writes that ``$p$ is incompatible with the set consisting
of $p\rightarrow q$ and $\thicksim q$, but not with either individually.''\cite{Brandom2008}
(123) I believe that after the arrow should be the set and that the
implication does not belong to the set. The set can be expressed as
a conjunctions of claims:
\begin{description}
\item [{(1)}] x is a blackberry is \emph{incompatible} with x is red $\wedge$
x is ripe
\item [{(2)}] y is a cherry is $\lnot$\emph{incompatible} with y is red
$\wedge$ y is ripe
\end{description}
Intuitively it could also be expressed in the following way: 
\begin{description}
\item [{(3)}] $p\wedge q\rightarrow\lnot r$ If x is a red and a blackberry,
then it is not ripe.
\end{description}
Problematic is that it would also be true in the case that the antecedent
is false. 

There are other sets of sentences that are incompatible and the problem
is here, that at the level of formal logic and universal statements,
it leads to inconsistencies, that can not be solved on the formal
level. The general statement that all birds fly has its problems with
species that belong to the genus bird, but do not fly, like e.g. penguins. 
\begin{description}
\item [{(4)}] x is a dove is $\lnot$\emph{inccompatible} with x is a bird
$\wedge$ x flies
\item [{(5)}] y is a penguin is \emph{incompatible} with y is a bird $\wedge$
y flies 
\end{description}
The statement in (5) on the left side is not incompatible with all
sentences, but just with one. If the sets are interpreted in a probabilistic
way then a weight is assigned to them according to how many sentences
are compatible or incompatible. Of course, a lot of logical fine-tuning
has to be done in order to understand the nature of concepts. The
expressive logical tools to do so need to be examined and the analytic
framework that contains them. The introduction of modal vocabulary
can help here.

\section{Modal Vocabulary}

Possible worlds are sometimes interpreted as existing parallel worlds
(David Lewis) or as worlds that are accesible from the actual world
and they represent different states of the actual world, which can
be expressed by counterfactuals (Saul Kripke).\cite{Kripke1963}\footnote{There is a widely used example that shows how a counterfactual statement
expresses the unlikelihood of a possible world and this makes it at
least seem difficult to take counterfactuals as adequate expressive
tools. There is a difference between: ``If Shakespeare did not write
Hamlet, someone else did.'' and ``If Shakespeare had not written
Hamlet, someone else would have.'' The counterfactual expresses in
way an idea that at least one person might have written the same piece
of art word by word. The example is from Jonathan Bennett.\cite{Bennett1988}
For an analysis of conditionals and the difference between indicative
and subjunctive conditionals see von Fintel (2012). The discussion
of possible worlds revolves around the similarity of the worlds and
the possible worlds are placed in the antecedent.\cite{VonFintel2012}
This is different to the approach that is proposed here.} Possible worlds are logically consistent within themselves and to
a certain degree the likelihood of the weights (predicates) can be
adjusted. But to start just with logical consistency, will never get
you to what is actually the case. All possible worlds seem to be equally
likely. That is why I suggest that the actual world should be represented
by a proposition that serves as the antecedent of a conditional. The
antecedent identifies or picks out the object that is classified in
the consequent.
\begin{description}
\item [{(6)}] $p\rightarrow q\wedge\lnot r$ If x is a blackberry, then
x is red and x is not ripe.
\end{description}
In the conditional the antecedent expresses a sufficient condition
for the consequent, while the consequent expresses a necessary condition.
Another example is: If Nellie is an elephant, she has a trunk. Being
an elephant is a sufficient condition of her to have a trunk and having
a trunk is a necesarry condition that she is an elephant.\cite{Brennan2017}
But the consequent expresses possible or dispositional properties.\footnote{Thomas Zoglauer writes that they express natural laws, but that is
not correct. Dispositional properties are not natural laws. They have
a different logical form.\cite{Zoglauer2002} (41)}
\begin{description}
\item [{(7)}] $p\rightarrow\diamondsuit(q\wedge\lnot r)$ If x is a blackberry,
then it is possible that x is red and x is not ripe.
\end{description}
The antecedent expresses here the starting point or the condition
that forms the possible world (consequent). It is a set of propositions
or properties which are possible, but also necessarily possible.\footnote{One axiom of the modal system \textbf{S5} is that something that is
possible is necessarily possible ($\diamondsuit p\rightarrow\boxempty\diamondsuit p$).
The necessary conditionality is expressed by the role the antecedent
and the consequent play in the truth values of the conditional. Maybe
the axiom of \textbf{S5} stems actually from the logical structure
of the implication.} It gives room for updating the properties, i.e. to change the proposition,
but also to regard them as necessary properties of the object. The
consequent expresses as a conjunction conceptual relations between
sets of propositions. I would call it the intensional part and the
possible world. Possible worlds express conceptual relations and not
ontological worlds or other realities. They are models of conceptual
relations. Even if the antecedent is not realized they are true, but
without the antecedent the relation has no context. It makes no sense
to say that something is possibly red and not ripe \textendash{} but
what exactly? \textendash{} We are talking about blackberries and
not about cherries.

\section{Limits of Labeling }

The underlying conceptual problem is that concepts, at least non-analytic
concepts, have only necessary characteristics and not necessary and
sufficient characteristics. For analytic concepts it might be possible
to have a list of necessary features of the concept and together they
are sufficient to grasp or \emph{describe} the concept. We do not
have an exhaustive list of features that constitute a concept. To
every characteristic another can be added: e.g. that elephants have
trunks, have four legs, are grey. Every feature is necessary, but
not sufficient to describe the species elephant. What can be selected
are only necessary features, when we label or cluster. Therefore,
conceptually there is already a limitation involved. \textendash{}
I believe that this limitation lies in the underlying logical structure
of concept use in classifying or labeling objects. (In the end, the
properties are only possible properties, but necessarily possible.)

This ideas can be linked to labeling in machine learning. A learning
algorithm can be supervised or unsupervised. Supervised learning means
that the data engineer knows what is ``correct'' or ``incorrect'',
it is labeled like this. In unsupervised learning clusters are build
by examples. John Guttag, in his book \emph{Introduction to Computation
and Programming Using Python}, uses as an example how to classify
animals as reptiles. Of course, we already know which animal is a
reptile and which is not. We use some examples to extract features
that, we suppose, are necessary conditions, like: is cold-blooded,
has scales, lays eggs, is poisonous, has no legs. These features are
extracted from the examples: cobra and rattlesnake. If we add boa
constrictor (to the training data), then we see that laying eggs and
poisonous are not necessary features. Unfortunately, a salmon would
now also be classified as a reptile and an alligator would not count
as an reptile. So, the feature has no legs has to be discarded as
irrelevant, too. (Or, that a reptile has either no legs or four legs.)\cite{Guttag2017}
(375-377) It depends here not only on the fact that we know which
animal is a reptile, but also on what is a necessary condition for
being a reptile.
\begin{quote}
``Does this mean that we should give up because all of the available
features are mere noise? No. In this case, the features \emph{scales}
and \emph{cold-blooded} are necessary conditions for being a reptile,
but not sufficient conditions. The rule that an animal is a reptile
if it has scales and is cold-blooded will not yield any false negatives,
i.e., any animal classified as non-reptile will indeed not be a reptile.
However, it will yield some false positives, i.e., some of the animals
classified as reptiles will not be reptiles.''\cite{Guttag2017}
(377)
\end{quote}
I think that he is correct that they are only necessary conditions,
but the necessity is based on a stipulation by us. \textendash{} Observable
and perceivable properties, like to have scales, are only conditional
generalities that are embedded in a context or situation or bound
to a singular thing. They are empirically verified.\footnote{This is the idea of abduction from Peirce. We stipulate a hypothesis
and see if it applies to a case, but it has to be undergirded empirically.} The stipulations are nor arbitrary, because the are undergirded empirically.
One can now count on them and calculate with them. The classification
encodes knowledge that can be used to ``calculate'' with it.

\section{Concept Use}

The consequent of the conditional can then be used as the antecedent
or premise for another conditional. 
\begin{description}
\item [{(8)}] $\diamondsuit\left(q\wedge\lnot r\right)\rightarrow\diamondsuit s$
\end{description}
If it is red and not ripe, then it is not eat\emph{able}. \textendash{}
The formalization of a dispositional predicate is much more difficult.
Some use conditionals and others use counterfactuals, but this would
lead us to another topic. Brandom writes that in 
\begin{quote}
``the most general sense, one classifies something simply by responding
to it differentially. Stimuli are grouped into kinds by the response-kinds
they tend to elicit. In this sense, a chunk of iron classifies its
environments into kinds by rusting in some of them and not others,
increasing or decreasing its temperature, shattering or remaining
intact. As is evident from this example, if classifying is just exercising
a reliable differential responsive disposition, it is a ubiquitous
feature of the inanimate world.''\cite{Brandom2009} (200/201)
\end{quote}
An algorithm would be a ``good'' algorithm, if it classifies the
objects reliably. Problematic is for Brandom that ``{[}c{]}lassification
as the exercise of reliable differential responsive dispositions (however
acquired) is not by itself yet a good candidate for \emph{conceptual}
classification, in the basic sense in which applying a concept to
something is \emph{describing} it.''\cite{Brandom2009} (202) He
builds his epistemological approach on Sellars, who states that in
order to describe objects we need to ``locate these objects in a
space of implications'' and therefore we do not ``merely label''
them\footnote{``It is only because the expressions in terms of which we describe
objects, even such basic expressions as words for the perceptible
characteristics of molar objects, locate these objects in a space
of implications, that they describe at all, rather than merely label.''\cite{Sellars1957}
(section 108)}, or merely classify them. It means not just to know in which situation
to respond reliably, but also to know the consequences of the application\footnote{``To learn what they\emph{ mean} is to learn, for instance, that
the owner put a red label on boxes to be discarded, green on those
to be retained, and yellow on those that needed further sorting and
decision. Once I know what \emph{follows} from affixing one rather
than another label, I can understand them not as \emph{mere} labels,
but as \emph{descriptions }of the boxes to which they are applied.
Description is classification with \emph{consequences}, either immediately
practical (\textquoteleft to be discarded/examined/kept') or for further
classifications.''\cite{Brandom2009} (203)} and to understand, when propositions or sets of propositions are
used as antecedents or consequents in conditionals or implications.
Canonized knowledge, which is based on empirical knowledge, can serve
as a premise of a conditional. It is a necessary and general knowledge
(\emph{a posteriori}) \textendash{} a possibility in the ``space
of implications'' that can be used to ``calculate'' with it or
to count on it. This means that it is reliable knowledge.

\emph{Space of implications:} The classification is: $p\rightarrow\diamondsuit(q\wedge\lnot r)$
(If x is a blackberry, then it is possible that x is red and x is
not ripe.) and the description is: $\diamondsuit\left(q\wedge\lnot r\right)\rightarrow\diamondsuit s$
(If it is red and not ripe, then it is not eatable.) One has to understand
which role the (sets of) propositions play (being premises or consequences)
in order to \textquotedblleft move\textquotedblright{} competently
in the web of inferences. That means also that there are different
kinds of concept use that only make sense and have meaning within
the ``space of implications''. 

The space of implications is a space of conceptual possibilities and
has to be elaborated. Which kind of reasons are good reasons in the
sense that they are employed in the right language game. The adequacy
of concept use and its location in the right language game are crucial
to distinguish between different kinds of reasons. This means that
statements are embedded in a logical space of conditions. There are
different kinds of conditions for statements: they can represent reasons,
conceptual relations, causes, or also motivational states that explain
actions. Inferring from the statements to its conditions is part of
abductive reasoning. 

\bibliographystyle{plain}
\bibliography{BibliographyLogicalUnderstanding}

\end{document}